\documentclass{llncs}

\usepackage{epsfig}
\usepackage{subfigure}
\usepackage{calc}
\usepackage{multicol}

\usepackage{listings}
 \usepackage{graphicx}
 \usepackage{tabulary}

\title{A tool for visualizing the execution of programs and stack traces especially suited for novice programmers}

\author{Stanislav Litvinov, Marat Mingazov, Vladislav Myachikov, \\ Vladimir Ivanov, Yuliya Palamarchuk, \\ Pavel Sozonov and Giancarlo Succi}
\date{}

\institute{Innopolis University \\ Innopolis, Russia}

\begin{document}

\maketitle


\abstract{Software engineering education and training has obstacles caused by a lack of basic knowledge about a process of program execution.
The article is devoted to the development of special tools that help to visualize the process. We analyze existing tools and propose a new approach to stack and heap visualization. The solution is able to overcome major drawbacks of existing tools and suites well for analysis of programs written in Java and C/C++.
\newline 

}



\section{Introduction}
Software engineering education and training has many obstacles. Failures of students are quite frequent even at introductory level programming courses; failure rate is approximately 33\% \cite{bennedsen2007failure,Pedrycz:2005,watson2014failure}. Apparently, students and novice programmers struggle with very basic concepts of programming, such as behavior of a program in run-time. Early failures in studying may dramatically decrease one's motivation to become a programmer \cite{Succi:C190.2008,oromachallenges,succi2001preliminary,Dragoni:2017}. 
Hence, this is a serious issue for education institutions, if students do not have a clear idea of how a program is executed.


One possible way to resolve the issue involves special tools for visualization the process of program execution. This requires a clear and easy to use visualization of stack and heap memory.
Despite many available solutions, very few existing tools are widespread \cite{sorva2013review}.  

In this paper we propose a new approach to visualization of stack and heap memory of a running program. The approach is aimed at building a clear understanding of program's memory organization. The visualization reflects a model of memory which suits to describe programs written on Java and C/C++.

Section \ref{S:Background} describes the state of the art. Section \ref{S:Solution} presents major parts of the solution: architecture, user interface, and visualization model. 
Section \ref{S:Conclusion} summarizes results and outlines the future work.

\section{Background} \label{S:Background}

\subsection{Modeling memory to support the learning process}
Each program in Java and C/C++ usually contains three separate segments: a program area, a stack, and a heap. The program area is where the code is located and it is used to access the instructions to execute.
The stack is used by processes or threads as a storage of arguments and local variables. This memory cannot be used if the data size is unknown ~\cite{Doe2016Misc,yurichev2013c,CsUmdEdu2003Misc}. 
The heap is more flexible, because size of data which will be stored in the heap can be determined at any moment, including run-time. However, usage of heap imposes significant overhead, since it needs additional processor instructions and memory space for storing memory allocation metadata \cite{Doe2016Misc,yurichev2013c}.

Usually, novice programmers and students misunderstand memory organization of a running program \cite{Safina:2016}. Typical learning challenges that novice programmers are facing with have been summarized in \cite{sorva2013review}:
(i) treating a program as a run-time process, not only a piece of code;
(ii) understanding of a computer working process;
(iii) revealing implicit programming constructs (e.g., pointers, references);
(iv) misunderstanding of program execution sequence and tracing.

Notably, that a standard debugger cannot help novice programmers due to its limited usability, not it contains useful metrics for defect detection \cite{Succi:C191.2008}. A debugger requires a knowledge of memory organization of a program that a novice typically does not have yet. Moreover, most debuggers do not provide any explanations or hints ~\cite{sorva2013review}. 

\subsection{Overview of visualization tools}
Most currently used programming languages for novice programmers are C/C++ and Java and existing visualization tools for them either emphasize their imperative features, or (for C++ and Java) the object oriented features (\cite{Succi:C124.2004,Succi:C229.2011,bennedsen2010bluej,Succi:C236.2011,Pedrycz:2015}).  Considering Java and C/C++ is also particularly relevant, since these languages are in the top 5 most popular languages \cite{bissyande2013popularity}.
Needless to say that this implies that this study focuses on compiled languages rather than on scripting languages.


\begin{table*}[!h]
\caption{A survey of tools for visualization of a program execution.}
\label{T:solutions2} 

\centering
\begin{tabulary}{\textwidth}{|C|L|}

\hline
   Title  & Description \\ \hline
   
 DYVISE \cite{reiss2009visualizing}     & a standalone application for analysis of memory leaks, inefficient use of memory, unexpected changes in memory, etc. \\ \hline
 
  JIVE  \cite{lessa2010jive}     & a plugin for Eclipse IDE; shows call history, method calls and object context; supports searching and stepping. \\ \hline
  
Trace \cite{alsallakh2012visual}    & a plugin for Eclipse IDE; a timeline represented as line chart with breakpoints on the line. 
\\ \hline

EXTRAVIS \cite{cornelissen2008execution}   & a prototype presents a program as an element in the circle with lines that represent relationships between classes and packages; and uses sequence diagram to overview events. \\ \hline

Memview \cite{gries2005memview}    &  an extension to the DrJava IDE; depicts call stack, static objects in heap, normal objects in heap in distinct boxes. \\ \hline
 
CoffeeDregs \cite{huizing2012visualization}  &  supports multithreading, but it is mostly a teaching tool for object-oriented study. \\ \hline
  
JaVis  \cite{mehner2002javis} &   a UML-based application; uses sequence diagram for time line and collaboration diagram for deadlock detection.   \\ \hline

JAVAVIS \cite{oechsle2002javavis} & it supports multithreading; shows stack call; uses a sequence diagram for a time-line and parallel threads. \\ \hline
 
EVizor \cite{moons2013design} & is a plugin for Netbeans IDE. Advantage of this application is textual tips for the user with explanations.  \\ \hline
  
 JavaTool \cite{mota2008javatool}  &  is a plugin for Moodle. It can be useful for small programs. Debugging occurs in the browser.  \\ \hline
 

Labster \cite{juett2016using} & a web-based system for visualization of memory representation and expressions evaluation.   \\ \hline

Project S \cite{deitz2016video} & a tool has graphical interface based on ``Space Invaders''. The aliens are variables. Each variable has a text label.    \\ \hline

Web-based tutor \cite{kumar2009data} & a tool depicts memory representation of C++ code: global variables, stack and heap. In addition, this tool has a detailed explanation related to a current line.  \\ \hline

VIP \cite{virtanen2005vip}  & a tool explains how pointers work in C++ and demonstrates the process of expression evaluation.  \\ \hline
 
Bradman  \cite{smith1995reinforcing}  &  an extended debugger which explains execution of each statement.  \\ \hline

Teaching Machine \cite{bruce2000lifting}   & a tool shows a stack, a heap and static memory. Special table includes a list of all variables: type, name and value.   \\ \hline
   
  jGRASP  \cite{cross2004jgrasp}  & an IDE. It can show visualization of data structures, objects, instance variables.  \\ \hline


\end{tabulary}
\end{table*}


In Table \ref{T:solutions2} we summarize some of the most prominent tools for visualizing execution of programs.
Most of the tools in Table \ref{T:solutions2} are research prototypes. Very few of them are still in active use. A limited number of tools have been used outside the place of its origin. However, in order to be effective in educational process a visualization tool should have high level of engagement \cite{juett2016using,moreno2004visualizing,Succi:C60.1999}.
To overcome learning challenges a visualization of stack and heap memory should be easy to use.

In general, these tools have the following advantages: (a) most of the applications have timeline or/and forward/reverse stepping, (b) one system has a flexible search mechanism which is able to work with many parameters (e.g., variable name, returned value, etc.)~\cite{lessa2010jive},  (c) some of the tools show multithreading and deadlocks, and  (d) textual explanation is very useful ~\cite{smith1995reinforcing}.

On the other side, we can identify the following, quite generalized, drawbacks: (a) only a few tools show separate heap and stack,  (b) not all of applications are convenient for beginners, (c) only two solutions work in Eclipse IDE, and (d) not all tools allow to use your own code.

\section{A novel approach to stack and heap visualization} \label{S:Solution}
We have devised a new approach for visualizing program execution process. The main element of the visualization is the stack trace, which makes memory organization of C/C++ and Java programs explicit\footnote{The source code of the prototype is available at https://github.com/MaratMingazov/CMemvit}.

We have implemented prototype to experiment our approach in three different environments:  (1)  Eclipse IDE for C++,  (2) Eclipse IDE for Java, and (3) IntelliJ IDEA for Java.

\subsection{System architecture of the prototypes} \label{S:Architecture}
\begin{figure}[!h]
   \centering
    {\epsfig{file = 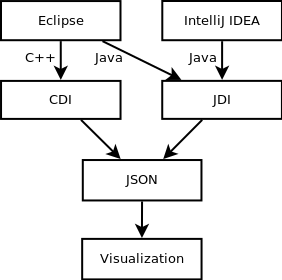, width = 5.5cm}}
   \caption{Application architecture}
   \label{fig:Architecture}
  \end{figure}

Figure~\ref{fig:Architecture} summarizes the general architecture of our prototypes. In the Java-based prototypes Eclipse and IntelliJ IDEA interact with Java Debugging Interface, and in the C/C++ prototypes Eclipse interacts with C/C++ Debugging Interface. 

Java Debugging Interface (JDI) is a high-level Java-based interface, which is directly used in debugger applications. JDI provides access to Java threads, virtual machine’s state, Class, Array, Interface, and primitive types ~\cite{Oracle2016Misc}.

C/C++ Debugging Interface (CDI) is a useful Java-based interface to custom debuggers in Eclipse environment. CDI can work with full-featured debugger provided by a development environment tooling (e.g., C/C++ Development Tooling (CDT)), or external debuggers (e.g., GDB). 
Eclipse plugins can interact with a debugger, and use all features of CDT environment, such as code-stepping, watchpoints, breakpoints, register contents, memory contents, variable views, signals, etc. Debugging results are shown in CDT Debugging perspective simultaneously ~\cite{scarpino2008cdt1,scarpino2008cdt2}.

At each execution step, when an event of changing process/thread state occurs, IDE plugin collects all the data about the actual process/thread state from CDI or JDI and generate a corresponding JSON object. The JSON object comprises several blocks: (1)  {\bf language}, specifying the programming language (Java / C++),
(2) {\bf threads}, only in Java, where there is at least one thread (main thread), and each thread contains its status and stack, (3) {\bf stack}, it is an independent block for C/C++, in case of Java it is inner block of "threads;" this element of JSON object contains information about stack frames and their content (function name, arguments, local variables, etc.), (4) {\bf heap}, including information about heap content,
(5) {\bf globalStaticVariables}, self defined, (6) {\bf lineNumber}, self defined.

The description of each variable includes several fields: name, type, value, address (for C++), identifier (for Java objects), etc. 
The JSON object is saved as a distinct file with unique name and timestamp. 

JSON files are sent to the visualization subsystem, which extracts data and builds a graphical representation connected with sequential execution steps. User can see current program state or state after any of previous steps. Work ~\cite{juett2016using} emphasizes the importance of possibility to use full-featured navigation in visualization system, i.e. not only to use standard stepping buttons (e.g., "step into", "step over", "step return"), but also to be able to return to previous states at any moment. 

The proposed architecture makes the development process scalable. Uniform representation of an intermediate JSON file precisely defines data to be extracted from debugger. A shared format optimizes development of the visualization subsystem. Indeed, instead of developing three different visualization modules we need to develop only one. In the future plugins for another IDE (or even other languages) can be easily developed and integrated in our architecture.

\subsection{User Interface} \label{S:UserInterface}
\begin{figure*}[h]
   \centering
    {\epsfig{file = 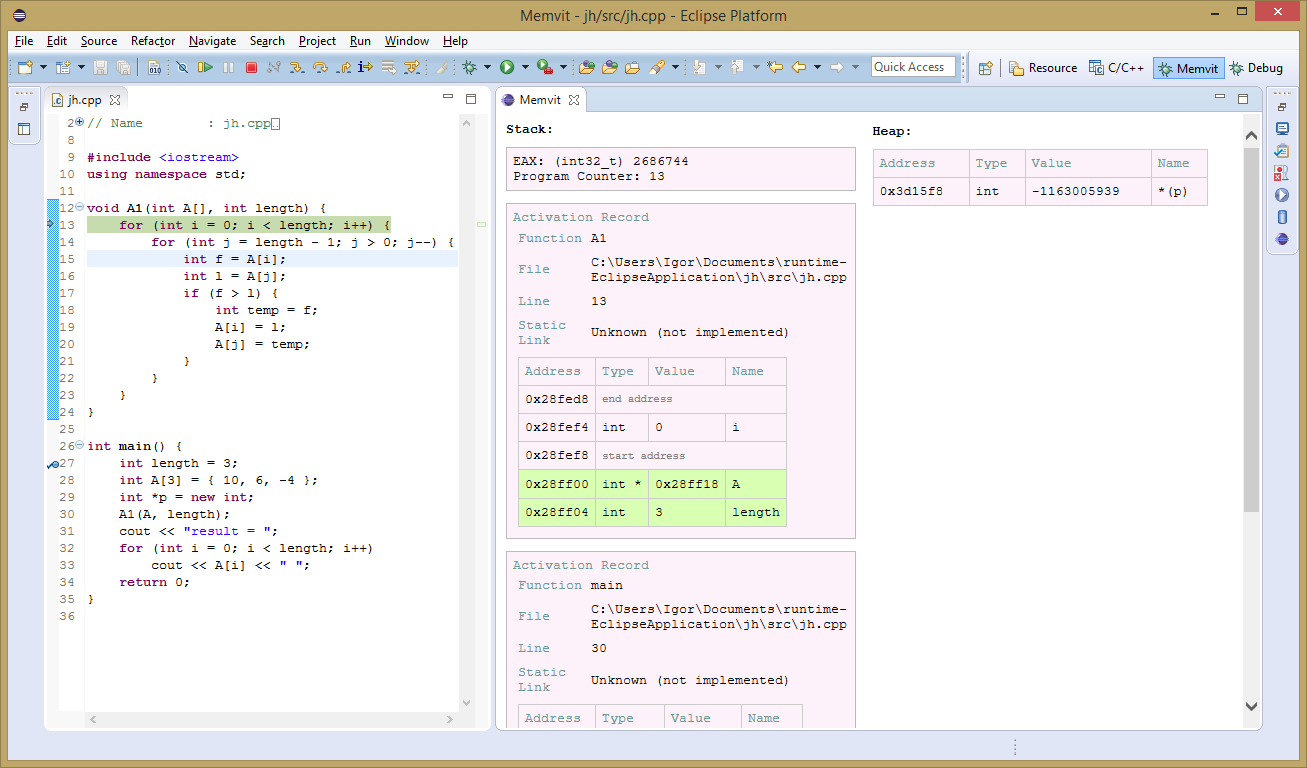, width = 12cm}}
   \caption{User Interface of the tool}
   \label{fig:user-interface}
  \end{figure*}

Technically, our application is an extension of an IDE (Java or Eclipse), which interacts to a built-in debugger. In a basic scenario user puts a breakpoint somewhere in the source code and then steps forward and backward, observing execution states. User interface of the tool is presented in Figure \ref{fig:user-interface}.

The user interface consists of: (a) {\it IDE standard window}; (b) {\it source code editor}, which also highlights current execution line, and its breakpoints managing functional; (c) {\it standard debug control buttons} (e.g., "step into", "step over", "step return") of IDE; (d) {\it view tab} with all visualization tables along with additional buttons for back-stepping, and visualization preferences. The visualization model of the application is illustrated with an example. To this end we use the following simple Java program. 

\begin{figure}[!h]
   \centering
    {\epsfig{file = 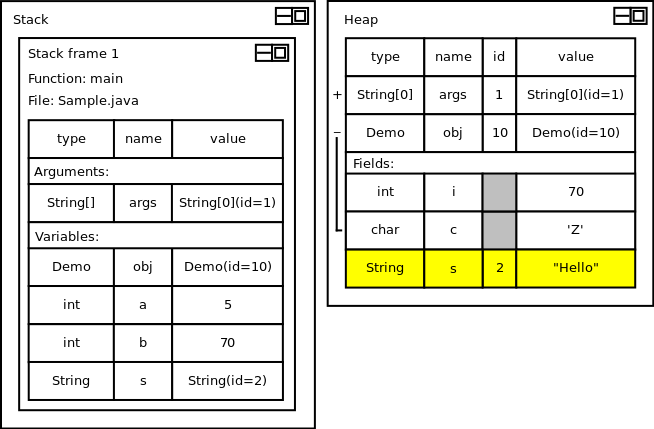, width = 6cm}}
   \caption{Heap and a stack states of a program (Listing \ref{JavaProgram})}
   \label{fig:VisualizationJava}
  \end{figure}

Figure \ref{fig:VisualizationJava} shows a heap and a stack state in the breakpoint. Inside the stack one can see arguments and local variables. Description of variables includes the following fields: type, name and value. For simplicity, names of standard classes are shown without prefix {\tt java.lang} (e.g., we show {\tt String} instead of {\tt java.lang.String}). In addition, we show only user's objects in the heap (only those objects, which have reference to them in the stack). Otherwise, the heap visualization might be littered with numerous system objects.
C++ visualization has a very similar structure, but it also contains global/static variables block, memory addresses of local variables, and memory addresses of heap objects instead of identifiers.

\begin{lstlisting}[language=Java, caption={Example of a Java program visualized in Fig. \ref{fig:VisualizationJava}}, label=JavaProgram]
public class Sample {
    public static void 
      main(String[] args) {
        Demo obj = new Demo();
        obj.i = 70;
        obj.c = 'Z';
        int a = 5;
        int b = obj.i;
        String s = "Hello";
   }//<-- current execution point
}
class Demo {
    int i;
    char c;
}
\end{lstlisting}

During the execution process each new stack frame appears on the top of the stack, and old frames move down. When a function finishes its execution, the stack frame of the function is removed from the stack. New heap objects or static variables appear on the top of the heap and static/global memory areas. Thus, even if user will work with a large program, visual representation will grow only vertically. This means that our approach allows users to effectively observe all the information about the program state using scrolling, and to have recent data always on the top. In addition, a user can customize the application. There is a possibility to automatically minimize all stack frames excluding the upper one and manually minimize or maximize any block (i.e., heap, global/static memory, stack or distinct stack frames). A variable or an object which was changed/created during the last step is highlighted. The field "name" is added into a table which represents the heap. This facilitates understanding the relationship between pointers/references in the stack and objects in the heap.

\section{Conclusion and future work} \label{S:Conclusion}
This article presents a new solution for visualization of program execution. Right now we have available only prototypes, but soon we are going to develop a working version and to test it.

The prototypes are plugins that allow us to monitor memory content of programs during execution step by step and that will be released with an Open Source license \cite{Succi:C144.2004}. It would be fruitful to pursue further research about including a timeline and textual explanations. If a timeline is shown as a sequence diagram, then we will be able to depict multithreading in our application. 
We have considered several advantages and disadvantages of existing visualization systems. Thus, we are going to gather some major advantages in one solution and eliminate flaws. So that novice programmers will obtain a powerful tool for understanding how programs execute and how memory is typically organized. 

\section*{Acknowledgements}
\noindent
We thank Innopolis University for supporting this research.

  \bibliographystyle{plain}

  {\small
  \bibliography{example}}
\end{document}